# Basis functions for solution of non-homogeneous wave equation

Sina Khorasani[*a], Farhad Karimi[b]
[a]School of Electrical Engineering, Sharif University of Technology, Tehran, Iran;
[b]Department of Electrical and Computer Engineering, University of Wisconsin, Madison, WI 53706

## ABSTRACT

In this note we extend the Differential Transfer Matrix Method (DTMM) for a second-order linear ordinary differential equation to the complex plane. This is achieved by separation of real and imaginary parts, and then forming a system of equations having a rank twice the size of the real-valued problem. The method discussed in this paper also successfully removes the problem of dealing with essential singularities, which was present in the earlier formulations. Then we simplify the result for real-valued problems and obtain a new set of basis functions, which may be used instead of the WKB solutions. These basis functions not only satisfy the initial conditions perfectly, but also, may approach the turning points without the divergent behavior, which is observed in WKB solutions. Finally, an analytical transformation in the form of a matrix exponential is presented for improving the accuracy of solutions.

**Keywords:** Quantum mechanics, Optics, Wave equation, Differential transfer matrix method,

## 1. FORMULATION

Suppose that the following problem, which models numerous optical and quantum mechanical problems, is given

$$y''(x) + f(x)y(x) = 0 \tag{1}$$

where $f(x) = g(x) + ih(x)$ and $y(x) = u(x) + iv(x)$ are complex-valued functions of $x$. For the moment, we assume that both functions are analytic in the complex plane. We can recast the above equation into the form

$$\frac{d}{dx}\begin{Bmatrix} u(x) \\ v(x) \\ u'(x) \\ v'(x) \end{Bmatrix} = \begin{bmatrix} [\mathbf{0}] & [\mathbf{1}] \\ [\mathbf{E}(x)] & [\mathbf{0}] \end{bmatrix} \begin{Bmatrix} u(x) \\ v(x) \\ u'(x) \\ v'(x) \end{Bmatrix} \tag{2}$$

in which

$$[\mathbf{0}] = \begin{bmatrix} 0 & 0 \\ 0 & 0 \end{bmatrix}$$
$$[\mathbf{1}] = \begin{bmatrix} 1 & 0 \\ 0 & 1 \end{bmatrix}$$
$$[\mathbf{E}(x)] = -\begin{bmatrix} g(x) & -h(x) \\ h(x) & g(x) \end{bmatrix} \tag{3}$$

Now we adopt the definitions

$$\{\mathbf{F}(x)\} = \begin{Bmatrix} u(x) \\ v(x) \\ u'(x) \\ v'(x) \end{Bmatrix} \tag{4}$$

---

[*] Email: khorasani@sina.sharif.edu; phone +98-912-304-3142; fax +98-21-6602-3261

$$[\mathbf{K}(x)] = \begin{bmatrix} [\mathbf{0}] & [\mathbf{1}] \\ [\mathbf{E}(x)] & [\mathbf{0}] \end{bmatrix}$$

(5)

to reach the DTMM-like system of equations

$$\frac{d}{dx}\{\mathbf{F}(x)\} = [\mathbf{K}(x)]\{\mathbf{F}(x)\}$$

(6)

subject to the initial conditions

$$\{\mathbf{F}(0)\} = \begin{Bmatrix} \Re f(0) \\ \Im f(0) \\ \Re f'(0) \\ \Im f'(0) \end{Bmatrix}$$

(7)

This system of equations can be readily solved *exactly* using the DTMM solution [1,2]

$$\{\mathbf{F}(x)\} = \mathbb{T}\exp\left[\int_0^x [\mathbf{K}(t)]dt\right]\{\mathbf{F}(0)\} = [\mathbf{Q}_{0 \to x}]\{\mathbf{F}(0)\}$$

(8)

The transfer matrices $[\mathbf{Q}_{p \to q}]$ satisfy the four basic properties [1]

$$[\mathbf{Q}_{p \to p}] = [\mathbf{1}] \quad \text{(Self-projection)}$$

$$[\mathbf{Q}_{p \to q}] = [\mathbf{Q}_{q \to p}]^{-1} \quad \text{(Inversion)}$$

$$|\mathbf{Q}_{p \to q}| = \exp(\text{tr}\{[\mathbf{K}_{p \to q}]\}) \quad \text{(Determinant)}$$

$$[\mathbf{Q}_{p \to r}] = [\mathbf{Q}_{q \to r}][\mathbf{Q}_{p \to q}] \quad \text{(Decomposition)}$$

(9)

Here, the Dyson's ordering operator $\mathbb{T}$ may be dropped at the expense of limited accuracy, while preserving the trace $\text{tr}\{[\mathbf{Q}_{0 \to x}]\}$ [3] along withthe first three properties above, to achieve the approximate, but explicit solution

$$\{\mathbf{F}(x)\} = \exp\left[\int_0^x [\mathbf{K}(t)]dt\right]\{\mathbf{F}(0)\} \cong [\mathbf{Q}_{0 \to x}]\{\mathbf{F}(0)\}$$

(10)

in which $\exp(\cdot)$ is the matrix exponentiation. We first define

$$[\mathbf{M}(x)] = \int_0^x [\mathbf{K}(t)]dt = \begin{bmatrix} [\mathbf{0}] & x[\mathbf{1}] \\ [\mathbf{B}(x)] & [\mathbf{0}] \end{bmatrix}$$

$$[\mathbf{B}(x)] = \int_0^x [\mathbf{E}(t)]dt = -\begin{bmatrix} G(x) & -H(x) \\ H(x) & G(x) \end{bmatrix}$$

(11)

Interestingly, the matrix exponentiation can be here simplified as follows

$$[\mathbf{M}(x)]^{2n} = x^n[\mathbf{B}(x)]^n \begin{bmatrix} [\mathbf{1}] & [\mathbf{0}] \\ [\mathbf{0}] & [\mathbf{1}] \end{bmatrix}$$

$$[\mathbf{M}(x)]^{2n+1} = x^n [\mathbf{B}(x)]^n \begin{bmatrix} [\mathbf{0}] & x[\mathbf{1}] \\ [\mathbf{B}(x)] & [\mathbf{0}] \end{bmatrix}$$

(12)

Hence

$$\begin{aligned}
\left[\mathbf{Q}_{0 \to x}\right] &= \exp[\mathbf{M}(x)] = \sum_{n=0}^{\infty} \frac{[\mathbf{M}(x)]^n}{n!} = \sum_{n=0}^{\infty} \frac{[\mathbf{M}(x)]^{2n}}{(2n)!} + \sum_{n=0}^{\infty} \frac{[\mathbf{M}(x)]^{2n+1}}{(2n+1)!} \\
&= \left( \sum_{n=0}^{\infty} x^n \frac{[\mathbf{B}(x)]^n}{(2n)!} \right) \begin{bmatrix} [\mathbf{1}] & [\mathbf{0}] \\ [\mathbf{0}] & [\mathbf{1}] \end{bmatrix} + \left( \sum_{n=0}^{\infty} x^n \frac{[\mathbf{B}(x)]^n}{(2n+1)!} \right) \begin{bmatrix} [\mathbf{0}] & x[\mathbf{1}] \\ [\mathbf{B}(x)] & [\mathbf{0}] \end{bmatrix} \\
&= \left( \sum_{n=0}^{\infty} \frac{[\mathbf{D}(x)]^{2n}}{(2n)!} \right) \begin{bmatrix} [\mathbf{1}] & [\mathbf{0}] \\ [\mathbf{0}] & [\mathbf{1}] \end{bmatrix} + \left( \sum_{n=0}^{\infty} \frac{[\mathbf{D}(x)]^{2n+1}}{(2n+1)!} \right) \begin{bmatrix} [\mathbf{0}] & x[\mathbf{D}(x)]^{-1} \\ \frac{1}{x}[\mathbf{D}(x)] & [\mathbf{0}] \end{bmatrix}
\end{aligned}$$

(13)

Here, we have assumed that $[\mathbf{D}(x)]$ is a matrix root of $[\mathbf{B}(x)]$ in such a way that (see Appendix A)

$$x[\mathbf{B}(x)] = [\mathbf{D}(x)]^2$$

$$\left[\mathbf{Q}_{0 \to x}\right] = \cosh[\mathbf{D}(x)] \begin{bmatrix} [\mathbf{1}] & [\mathbf{0}] \\ [\mathbf{0}] & [\mathbf{1}] \end{bmatrix} + \sinh[\mathbf{D}(x)] \begin{bmatrix} [\mathbf{0}] & x[\mathbf{D}(x)]^{-1} \\ \frac{1}{x}[\mathbf{D}(x)] & [\mathbf{0}] \end{bmatrix}$$

(14)

We here notice that the matrix hyperbolic functions $\sinh(\cdot)$ and $\cosh(\cdot)$ are defined according to their Taylor expansions, similar to the matrix exponentiation $\exp(\cdot)$. Evaluation of these matrix functions are discussed in Appendix B. Now, we assume that $\left[\mathbf{Q}_{0 \to x}\right] = \left[q_{0 \to x}^{ij}\right]$. The final solution is then simply

$$y(x) = [q_{0 \to x}^{11} u(0) + q_{0 \to x}^{12} v(0) + q_{0 \to x}^{13} u'(0) + q_{0 \to x}^{14} v'(0)] + i[q_{0 \to x}^{21} u(0) + q_{0 \to x}^{22} v(0) + q_{0 \to x}^{23} u'(0) + q_{0 \to x}^{24} v'(0)]$$

(15)

We rename

$$\begin{aligned}
[\mathbf{C}(x)] &= \cosh\left[\sqrt{x}\mathbf{D}(x)\right] = \left[C_{ij}(x)\right] \\
[\mathbf{S}(x)] &= x[\mathbf{D}(x)]^{-1} \sinh[\mathbf{D}(x)] = \left[S_{ij}(x)\right]
\end{aligned}$$

(16)

Then, the solution $y(x) = u(x) + iv(x)$ is now already known with

$$\begin{aligned}
u(x) &= C_{11}(x)u(0) + C_{12}(x)v(0) + S_{11}(x)u'(0) + S_{12}(x)v'(0) \\
v(x) &= C_{21}(x)u(0) + C_{22}(x)v(0) + S_{21}(x)u'(0) + S_{22}(x)v'(0)
\end{aligned}$$

(17)

The derivative $y'(x) = u'(x) + iv'(x)$ will be evaluated similarly

$$\begin{aligned}
u'(x) &= T_{11}(x)u(0) + T_{12}(x)v(0) + C_{11}(x)u'(0) + C_{12}(x)v'(0) \\
v'(x) &= T_{21}(x)u(0) + T_{22}(x)v(0) + C_{21}(x)u'(0) + C_{22}(x)v'(0)
\end{aligned}$$

(18)

where

$$[\mathbf{T}(x)] = \frac{1}{x}[\mathbf{B}(x)][\mathbf{S}(x)] = \left[T_{ij}(x)\right]$$

(19)

## 2. REDUCTION TO THE REAL-AXIS

As an application, we demonstrate how to use this method when dealing with real-valued problems having the type

$$u''(x) + g(x)u(x) = 0$$

(20)

The previous formulations of DTMM [1,4] would fail at zeros of $g(x)$. The method of Airy functions [5] and generalized DTMM [2] are able to respectively pass over and remove certain types of these singularities. This extended method, however, effectively works out such singular points, since the arising singularities are now located at zeros of $\int_0^x g(t)dt$ instead of $g(x)$, and furthermore the only singular expression corresponds to $[\mathbf{S}(x)]$. But it has been is shown in Appendix C that this singularity could be easily smoothed out. Now, for application to the real-plane, it is simply sufficient to take $v(0) = v'(0) = 0$, to obtain the approximate DTMM solution

$$u(x) = C_{11}(x)u(0) + S_{11}(x)u'(0)$$
(21)

while $x$ is strictly restricted to the real axis. These relations can be further simplified noting that $h(x) = 0$ and therefore $H(x) = 0$. But according to Appendix A, $a = 0$ and $b = \sqrt{xG(x)}$. Hence, we obtain after some algebra

$$C_{11}(x) = \cos\sqrt{x \int_0^x g(t)dt}$$

$$S_{11}(x) = x\,\mathrm{sinc}\sqrt{x \int_0^x g(t)dt}$$
(22)

where $\mathrm{sinc}(x) = \frac{1}{x}\sin(x)$. It can be easily verified that these functions must be respectively even with $C_{11}(x) = C_{11}(-x)$ and odd with $S_{11}(x) = -S_{11}(-x)$, if $g(x) = g(-x)$ is also an even function of $x$. Furthermore, the above given solution is accurate in the neighborhood of $x = 0$. For obtaining a solution accurate close to $x = \alpha$, we may simply shift the origin to obtain

$$u(x) = \Psi_1(x;\alpha)u(\alpha) + \Psi_2(x;\alpha)u'(\alpha)$$

$$\Psi_1(x;\alpha) = \cos\sqrt{(x-\alpha)\int_\alpha^x g(t)dt}$$

$$\Psi_2(x;\alpha) = (x-\alpha)\,\mathrm{sinc}\sqrt{(x-\alpha)\int_\alpha^x g(t)dt}$$
(23)

## 3. BASIS FUNCTIONS

In order to make a comparison to other approximate methods, we first define $g(x) = k^2(x)$. The well-known WKB basis functions are given by

$$U_1(x) = \frac{1}{\sqrt{k(x)}}\cos\left(\int_0^x k(t)dt\right)$$

$$U_2(x) = \frac{1}{\sqrt{k(x)}}\sin\left(\int_0^x k(t)dt\right)$$
(24)

In contrast, the method in this paper suggests the basis functions

$$\Psi_1(x) = \cos\sqrt{x\int_0^x k^2(t)dt}$$

$$\Psi_2(x) = x\,\text{sinc}\sqrt{x\int_0^x k^2(t)dt}$$

(25)

Now, we investigate the usefulness of these two new basis functions. For the case of $g = k^2$ being a constant, we get the answers

$$\Psi_1(x) = \cos(kx)$$
$$\Psi_2(x) = \frac{1}{k}\sin(kx)$$

(26)

which correspond to the exact solution with sinusoidal basis functions. WKB solutions also generate similarly exact solutions. While WKB basis functions are divergent near a turning point at which $k(x) = 0$, the presented set of basis functions clearly are not.

At the initial starting point, we may also observe that

$$\Psi_1(0) = 1$$
$$\Psi_2(0) = 0$$

(27)

and hence the initial condition at $u(0)$ is satisfied. After some considerable algebra (Appendix D), it may be shown that

$$\Psi_1'(0) = 0$$
$$\Psi_2'(0) = 1$$

(28)

Therefore, the initial condition at $u'(0)$ is also satisfied. It is worth mentioning that there is an improved basis founded on Airy functions, which is obtained using the method of asymptotic expansions [6]. These basis functions, however, are too complicated for analytical purposes and hand calculations. Furthermore, these basis functions need to be exchanged across the turning points for smooth solutions.

## 4. IMPROVED ACCURACY

Now, we return to the original problem

$$\{F(x)\}' = [K(x)]\{F(x)\}$$

(29)

having the exact and approximate solutions

$$\{F(x)\} = \mathbb{T}\exp[M(x)]\{F(0)\}$$
$$\{\Phi(x)\} = \exp[M(x)]\{F(0)\}$$
$$[M(x)] = \int_0^x [K(t)]dt$$

(30)

The exact and approximate transfer matrices are respectively given by

$$[\mathbf{Q}_{0 \to x}] = \mathbb{T}\exp[\mathbf{M}(x)]$$
$$[\mathbf{P}_{0 \to x}] = \exp[\mathbf{M}(x)]$$
(31)

In most cases, direct analytical evaluation of the exact transfer matrix $[\mathbf{Q}_{0 \to x}]$ turns out to be impossible. Here, we show how to construct a transformation which could project $[\mathbf{P}_{0 \to x}]$ unto $[\mathbf{Q}_{0 \to x}]$. Suppose that for any arbitrary $x$, we could define a transformation matrix $[\mathbf{W}(x)]$ such that

$$[\mathbf{Q}_{0 \to x}] = [\mathbf{W}(x)][\mathbf{P}_{0 \to x}]$$
(32)

Then by differentiating both sides with respect to $x$ we obtain

$$[\mathbf{K}(x)][\mathbf{Q}_{0 \to x}] = [\mathbf{W}(x)]'[\mathbf{P}_{0 \to x}] + [\mathbf{W}(x)][\mathbf{P}_{0 \to x}]'$$
(33)

Then we have

$$[\mathbf{W}(x)]' = ([\mathbf{K}(x)][\mathbf{W}(x)][\mathbf{P}_{0 \to x}] - [\mathbf{W}(x)][\mathbf{P}_{0 \to x}]')[\mathbf{P}_{0 \to x}]^{-1}$$
(34)

Noting that we must have $[\mathbf{W}(0)] = [\mathbf{1}]$, the above can be integrated as

$$[\mathbf{W}(x)] = [\mathbf{1}] + \int_0^x ([\mathbf{K}(t)][\mathbf{W}(t)] - [\mathbf{W}(t)][\mathbf{P}_{0 \to t}]'[\mathbf{P}_{0 \to t}]^{-1})dt$$
(35)

Clearly, the above equation can be recursively solved to obtain the successive iterations

$$[\mathbf{W}_0(x)] = [\mathbf{1}]$$
$$[\mathbf{W}_{n+1}(x)] = [\mathbf{1}] + \int_0^x ([\mathbf{K}(t)][\mathbf{W}_n(t)] - [\mathbf{W}_n(t)][\mathbf{P}_{0 \to t}]'[\mathbf{P}_{0 \to t}]^{-1})dt$$
(36)

This gives to a first-order correction

$$[\mathbf{W}(x)] \cong [\mathbf{1}] + \int_0^x ([\mathbf{K}(t)] - [\mathbf{P}_{0 \to t}]'[\mathbf{P}_{0 \to t}]^{-1})dt$$
(37)

The derivative $[\mathbf{P}_{0 \to x}]'$ can be expanded using Wilcox's expression [7] and Baker–Campbell–Hausdorff formula [8] as

$$\frac{d}{dx}[\mathbf{P}_{0 \to x}] = \int_0^1 \exp(\alpha[\mathbf{M}(x)])[\mathbf{K}(x)]\exp((1-\alpha)[\mathbf{M}(x)])d\alpha$$
$$= \left\{[\mathbf{K}(x)] + \frac{1}{2!}[[\mathbf{M}(x)],[\mathbf{K}(x)]] + \frac{1}{3!}[[\mathbf{M}(x)],[[\mathbf{M}(x)],[\mathbf{K}(x)]]] + \cdots\right\}[\mathbf{P}_{0 \to x}]$$
(38)

Another preferred method to solve the above differential equation approximately is to first let the commutator $[[\mathbf{K}(t)],[\mathbf{W}(t)]]$ vanish. Then we obtain

$$[\mathbf{W}(x)]' = [\mathbf{W}(x)]([\mathbf{K}(x)] - [\mathbf{P}_{0 \to x}]'[\mathbf{P}_{0 \to x}]^{-1})$$
(39)

Now by using the Wilcox expansion and retaining the first non-vanishing term we get

$$[\mathbf{W}(x)]' = -\frac{1}{2}[\mathbf{W}(x)][[\mathbf{M}(x)],[\mathbf{K}(x)]]$$
(40)

This, after dropping the ordering operator $\mathbb{T}$ again, allows the approximate solution

$$[\mathbf{W}(x)] = \exp\left\{-\frac{1}{2}\int_0^x [[\mathbf{M}(t)],[\mathbf{K}(t)]]dt\right\}$$

(41)

Hence, an improved solution is now given by

$$[\mathbf{Q}_{0\to x}] = \exp\left\{-\frac{1}{2}\int_0^x [[\mathbf{M}(t)],[\mathbf{K}(t)]]dt\right\}\exp[\mathbf{M}(x)]$$

$$\{\mathbf{F}(x)\} = \exp\left\{-\frac{1}{2}\int_0^x [[\mathbf{M}(t)],[\mathbf{K}(t)]]dt\right\}\exp[\mathbf{M}(x)]\{\mathbf{F}(0)\}$$

(42)

It is well-known that when the commutator $[[\mathbf{M}(t)],[\mathbf{K}(t)]]$ is zero [1,3], then dropping $\mathbb{T}$ should have no effect and hence $[\mathbf{P}_{0\to x}] = \exp[\mathbf{M}(x)]$ must be already exact. This fact is also justified by the above form since the correction exponent vanishes when this commutator is zero. There are several other known *sufficient* conditions for this to happen, which are discussed elsewhere [3,9].

It is not difficult to evaluate this correction term. As a matter of fact we have

$$[[\mathbf{M}(t)],[\mathbf{K}(t)]] = \begin{bmatrix} [\mathbf{0}] & t[\mathbf{1}] \\ [\mathbf{B}(t)] & [\mathbf{0}] \end{bmatrix}\begin{bmatrix} [\mathbf{0}] & [\mathbf{1}] \\ [\mathbf{E}(t)] & [\mathbf{0}] \end{bmatrix} - \begin{bmatrix} [\mathbf{0}] & [\mathbf{1}] \\ [\mathbf{E}(t)] & [\mathbf{0}] \end{bmatrix}\begin{bmatrix} [\mathbf{0}] & t[\mathbf{1}] \\ [\mathbf{B}(t)] & [\mathbf{0}] \end{bmatrix}$$
$$= \begin{bmatrix} t[\mathbf{E}(t)]-[\mathbf{B}(t)] & [\mathbf{0}] \\ [\mathbf{0}] & [\mathbf{B}(t)]-t[\mathbf{E}(t)] \end{bmatrix}$$

(43)

Then

$$\int_0^x [[\mathbf{M}(t)],[\mathbf{K}(t)]]dt = \begin{bmatrix} [\mathbf{J}(x)] & [\mathbf{0}] \\ [\mathbf{0}] & -[\mathbf{J}(x)] \end{bmatrix}$$

$$[\mathbf{J}(x)] = \int_0^x t[\mathbf{E}(t)] - [\mathbf{B}(t)]dt$$

(44)

Hence

$$[\mathbf{W}(x)] = \exp\left\{-\frac{1}{2}\begin{bmatrix} [\mathbf{J}(x)] & [\mathbf{0}] \\ [\mathbf{0}] & -[\mathbf{J}(x)] \end{bmatrix}\right\} = \begin{bmatrix} \exp\{-\frac{1}{2}[\mathbf{J}(x)]\} & [\mathbf{0}] \\ [\mathbf{0}] & \exp\{+\frac{1}{2}[\mathbf{J}(x)]\} \end{bmatrix}$$

(45)

For real-valued problems with

$$[\mathbf{E}(x)] = -g(x)[\mathbf{1}]$$
$$[\mathbf{B}(x)] = -G(x)[\mathbf{1}]$$

(46)

we get

$$[\mathbf{J}(x)] = \left(\int_0^x -tg(t) + G(t)dt\right)[\mathbf{1}] = \gamma(x)[\mathbf{1}]$$

(47)

Then

$$[\mathbf{W}(x)] = \begin{bmatrix} \exp\left\{-\frac{1}{2}\gamma(x)[\mathbf{1}]\right\} & [\mathbf{0}] \\ [\mathbf{0}] & \exp\left\{+\frac{1}{2}\gamma(x)[\mathbf{1}]\right\} \end{bmatrix} = \begin{bmatrix} \exp\left\{-\frac{1}{2}\gamma(x)\right\}[\mathbf{1}] & [\mathbf{0}] \\ [\mathbf{0}] & \exp\left\{+\frac{1}{2}\gamma(x)\right\}[\mathbf{1}] \end{bmatrix}$$
(48)

Hence, we obtain

$$[\mathbf{Q}_{0\to x}] = [\mathbf{W}(x)][\mathbf{P}_{0\to x}]$$
$$= \begin{bmatrix} \exp\left\{-\frac{1}{2}\gamma(x)\right\}[\mathbf{1}] & [\mathbf{0}] \\ [\mathbf{0}] & \exp\left\{+\frac{1}{2}\gamma(x)\right\}[\mathbf{1}] \end{bmatrix} \cosh[\mathbf{D}(x)] \begin{bmatrix} [\mathbf{1}] & [\mathbf{0}] \\ [\mathbf{0}] & [\mathbf{1}] \end{bmatrix}$$
$$+ \begin{bmatrix} \exp\left\{-\frac{1}{2}\gamma(x)\right\}[\mathbf{1}] & [\mathbf{0}] \\ [\mathbf{0}] & \exp\left\{+\frac{1}{2}\gamma(x)\right\}[\mathbf{1}] \end{bmatrix} \sinh[\mathbf{D}(x)] \begin{bmatrix} [\mathbf{0}] & x[\mathbf{D}(x)]^{-1} \\ \frac{1}{x}[\mathbf{D}(x)] & [\mathbf{0}] \end{bmatrix}$$
(49)

or

$$[\mathbf{Q}_{0\to x}] = \begin{bmatrix} \exp\left\{-\frac{1}{2}\gamma(x)\right\}[\mathbf{C}(x)] & \exp\left\{-\frac{1}{2}\gamma(x)\right\}[\mathbf{S}(x)] \\ \exp\left\{+\frac{1}{2}\gamma(x)\right\}[\mathbf{T}(x)] & \exp\left\{+\frac{1}{2}\gamma(x)\right\}[\mathbf{C}(x)] \end{bmatrix}$$
(50)

Here, we notice that for $g(x)$ and $\gamma(x)$ encompass identical even or odd symmetries. Finally, the above will result in the improved real-valued solution, after proper shifting the origin, expressed as

$$u(x) = \Psi_1(x;\alpha)u(\alpha) + \Psi_2(x;\alpha)u'(\alpha)$$
$$u'(x) = \Psi_3(x;\alpha)u(\alpha) + \Psi_4(x;\alpha)u'(\alpha)$$

$$\Psi_1(x;\alpha) = \exp\left\{-\frac{1}{2}\gamma(x;\alpha)\right\} \cos\sqrt{(x-\alpha)\int_\alpha^x g(t)dt}$$

$$\Psi_2(x;\alpha) = \exp\left\{-\frac{1}{2}\gamma(x;\alpha)\right\}(x-\alpha)\operatorname{sinc}\sqrt{(x-\alpha)\int_\alpha^x g(t)dt}$$

$$\Psi_3(x;\alpha) = -\exp\left\{+\frac{1}{2}\gamma(x;\alpha)\right\}\left(\int_\alpha^x g(t)dt\right)\operatorname{sinc}\sqrt{(x-\alpha)\int_\alpha^x g(t)dt}$$

$$\Psi_4(x;\alpha) = \exp\left\{+\frac{1}{2}\gamma(x;\alpha)\right\} \cos\sqrt{(x-\alpha)\int_\alpha^x g(t)dt}$$

$$\exp\left\{-\frac{1}{2}\gamma(x;\alpha)\right\} = \exp\left[-\frac{1}{2}\int_\alpha^x\left(-(t-\alpha)g(t) + \int_\alpha^t g(s)ds\right)dt\right]$$
(51)

# 5. ALGORITHM FOR ACCURATE SEMI-ANALYTICAL SOLUTIONS

The presented basis functions extend continuously and accurately in the neighborhood of the initial point with $x = \alpha$, while satisfying the initial conditions exactly. However, their accuracy start to diminish at farther points. This difficulty can be easily avoided by sectioning the solution domain into several sections $x \in U_n$ with $U_n = [x_n, x_{n+1}[ = [x_n, x_n + \Delta_n[$ and $\alpha = x_0$. These sections clearly need not to be equal. Here, we limit the discussion to real-valued problems, and the solution can then be recursively extended from $U_{n-1}$ to $U_n$ as

$$u(x \in U_n) = \Psi_1(x; x_n)u(x_n) + \Psi_2(x; x_n)u'(x_n)$$
$$u'(x \in U_n) = \Psi_3(x; x_n)u(x_n) + \Psi_1(x; x_n)u'(x_n)$$

(52)

where $u(x_n)$ and $u'(x_n)$ are taken and already known from $u(x \in U_{n-1})$, using the relevant expressions

$$u(x_n) = \Psi_1(x_n; x_{n-1})u(x_{n-1}) + \Psi_2(x_n; x_{n-1})u'(x_{n-1})$$
$$u'(x_n) = \Psi_3(x_n; x_{n-1})u(x_{n-1}) + \Psi_1(x_n; x_{n-1})u'(x_{n-1})$$

(53)

In the above equations, we have

$$\Psi_3(x; \alpha) = -\frac{1}{x - \alpha}\left(\int_\alpha^x g(t)dt\right)\Psi_1(x; \alpha)$$

(54)

Equivalently, using transfer matrix formalism we may concisely write

$$\{\mathbf{F}(x \in U_n)\} = [\mathbf{Q}_{x_n \to x}]\{\mathbf{F}(x_n)\}$$
$$\{\mathbf{F}(x_n)\} = [\mathbf{Q}_{x_{n-1} \to x_n}]\{\mathbf{F}(x_{n-1})\}$$

(55)

# 6. PERIODIC PROBLEMS

For those problems having periodicity, such as $f(x) = f(x + L)$, it is possible to use DTMM to obtain the Bloch eigenfunctions [2,4]. The Bloch eigenfunctions satisfy

$$y(x; \kappa) = \exp(ix\kappa)\Theta(x; \kappa)$$
$$\Theta(x; \kappa) = \Theta(x + L; \kappa)$$

(56)

in which $\kappa$ is the Bloch wavenumber. The governing differential equation for the so-called envelope functions $\Theta(x; \kappa)$ will be given by

$$\Theta''(x; \kappa) + 2i\kappa\Theta'(x; \kappa) + [f(x) - \kappa^2]\Theta(x; \kappa) = 0$$

(57)

Using the algorithm discussed in the last section, we may take on $\Delta_n = L$ to write

$$u(x + L) = \Psi_1(x + L; x)u(x) + \Psi_2(x + L; x)u'(x)$$
$$u'(x + L) = \Psi_3(x + L; x)u(x) + \Psi_1(x + L; x)u'(x)$$

(58)

or

$$\{\mathbf{F}(x + L)\} = [\mathbf{Q}_{x \to x+L}]\{\mathbf{F}(x)\}$$

(59)

Meanwhile, Bloch boundary conditions require

$$u(x + L) = \exp(i\kappa L)u(x)$$
$$u'(x + L) = \exp(i\kappa L)u'(x)$$

(60)

Therefore, we obtain the simple result for the Bloch wave number $\kappa$

$$\exp(i\kappa L) = \text{eig}[\mathbf{Q}_{x \to x+L}]$$

(61)

The Bloch wave number $\kappa$ must be in dependent of $x$ [2,4], but this is not exactly satisfied since $[\mathbf{Q}_{x \to x+L}]$ is not found exactly. But the correction exponent as introduction in Section 4 is a major correction to alleviate this deficiency, and hence $x$ may be, for instance, arbitrarily set to zero or $-\frac{1}{2}L$.

## 7. CONCLUSIONS

The main contribution of this paper is two-fold. Firstly, we have succeeded in presenting a new basis for solution of the non-homogeneous wave equation, instead of WKB solutions. These alternative basis functions provide non-divergent behavior near turning points and remain remarkably close to the analytic solution in the neighborhood of initial point. The initial conditions are here satisfied exactly. The second contribution is a general method for correction of solutions to an arbitrary order of accuracy. We successfully have evaluated the correction term to the first order. The results of this paper may find novel applications in optics and quantum mechanics, as well as control theory and system engineering.

## APPENDIX A. MATRIX SQUARE ROOT

The matrix square root of the $2 \times 2$ matrix $[\mathbf{B}]$ defined as
$$x[\mathbf{B}] = [\mathbf{D}]^2 \qquad (62)$$
can be evaluated relatively easy. We first note that
$$[\mathbf{B}] = \begin{bmatrix} -G & +H \\ -H & -G \end{bmatrix} \qquad (63)$$
Then, it is not difficult to see that $[\mathbf{D}]$ should also take a similar form as
$$[\mathbf{D}] = \begin{bmatrix} a & b \\ -b & a \end{bmatrix} \qquad (64)$$
Hence, we reach the system of equations
$$a^2 - b^2 = -Gx, \quad 2ab = Hx \qquad (65)$$
This can be rearranged into the quartic equation
$$a^4 + Gxa^2 - \frac{Hx}{4} = 0 \qquad (66)$$
having the solution
$$a = \pm\sqrt{x\frac{\sqrt{G^2 + H^2} - G}{2}}, \qquad b = \pm\sqrt{x\frac{\sqrt{G^2 + H^2} + G}{2}} \qquad (67)$$
Hence, we get the square roots
$$[\mathbf{D}] = \pm\sqrt{\frac{x}{2}} \begin{bmatrix} +\sqrt{\sqrt{G^2 + H^2} - G} & +\sqrt{\sqrt{G^2 + H^2} + G} \\ -\sqrt{\sqrt{G^2 + H^2} + G} & +\sqrt{\sqrt{G^2 + H^2} - G} \end{bmatrix} \qquad (68)$$
The choice of plus or minus sign in the present formulation of DTMM is immaterial.

## APPENDIX B. $2 \times 2$ MATRIX HYPERBOLIC FUNCTIONS

For a $2 \times 2$ matrixsuch as $[\mathbf{D}]$ with the form discussed in Appendix A, the exact matrix exponential can be found using the expression given in the reference [4]. Then, the exact hyperbolic functions can be found using
$$\cosh[\mathbf{D}] = \frac{\exp[+\mathbf{D}] + \exp[-\mathbf{D}]}{2}, \qquad \sinh[\mathbf{D}] = \frac{\exp[+\mathbf{D}] - \exp[-\mathbf{D}]}{2} \qquad (69)$$

We first construct the traceless matrix $[\mathbf{N}]$ as

$$[\mathbf{N}] = b\begin{bmatrix} 0 & 1 \\ -1 & 0 \end{bmatrix} = b[\bar{\mathbf{1}}]$$

(70)

Evidently, $[\bar{\mathbf{1}}]^2 = -[\mathbf{1}]$. Then, we have

$$[\mathbf{D}] = a[\mathbf{1}] + b[\bar{\mathbf{1}}]$$

(71)

Hence, we get

$$\exp[+\mathbf{D}] = \exp[a[\mathbf{1}] + b[\bar{\mathbf{1}}]] = e^a \exp[b[\bar{\mathbf{1}}]]$$

(72)

But we have

$$(b[\bar{\mathbf{1}}])^{2n} = (-1)^n b^{2n}[\mathbf{1}], \qquad (b[\bar{\mathbf{1}}])^{2n+1} = (-1)^n b^{2n+1}[\bar{\mathbf{1}}]$$

(73)

Therefore, we obtain

$$\exp[\mathbf{D}] = e^a \exp[b[\bar{\mathbf{1}}]] = e^a \sum_{n=0}^{\infty} \frac{(b[\bar{\mathbf{1}}])^n}{n!} = e^a \sum_{n=0}^{\infty} \frac{(b[\bar{\mathbf{1}}])^{2n}}{(2n)!} + e^a \sum_{n=0}^{\infty} \frac{(b[\bar{\mathbf{1}}])^{2n+1}}{(2n+1)!}$$

$$= e^a \sum_{n=0}^{\infty} \frac{(-1)^n b^{2n}}{(2n)!}[\mathbf{1}] + e^a \sum_{n=0}^{\infty} \frac{(-1)^n b^{2n+1}}{(2n+1)!}[\bar{\mathbf{1}}]$$

$$= e^a \cos(b)[\mathbf{1}] + e^a \sin(b)[\bar{\mathbf{1}}]$$

$$= e^a \begin{bmatrix} \cos(b) & \sin(b) \\ -\sin(b) & \cos(b) \end{bmatrix}$$

(74)

Therefore, we immediately obtain

$$\cosh[\mathbf{D}] = \frac{e^a}{2}\begin{bmatrix} \cos(b) & \sin(b) \\ -\sin(b) & \cos(b) \end{bmatrix} + \frac{e^{-a}}{2}\begin{bmatrix} \cos(b) & -\sin(b) \\ \sin(b) & \cos(b) \end{bmatrix}$$

$$\sinh[\mathbf{D}] = \frac{e^a}{2}\begin{bmatrix} \cos(b) & \sin(b) \\ -\sin(b) & \cos(b) \end{bmatrix} - \frac{e^{-a}}{2}\begin{bmatrix} \cos(b) & -\sin(b) \\ \sin(b) & \cos(b) \end{bmatrix}$$

(75)

These can be further simplified as

$$\cosh[\mathbf{D}] = \begin{bmatrix} \cosh(a)\cos(b) & \sinh(a)\sin(b) \\ -\sinh(a)\sin(b) & \cosh(a)\cos(b) \end{bmatrix}$$

$$\sinh[\mathbf{D}] = \begin{bmatrix} \sinh(a)\cos(b) & \cosh(a)\sin(b) \\ -\cosh(a)\sin(b) & \sinh(a)\cos(b) \end{bmatrix}$$

(76)

## APPENDIX C. DEALING WITH SINGULARITIES

The determinant of $[\mathbf{D}]$ as calculated in Appendix A is given by

$$|\mathbf{D}| = x\sqrt{G^2 + H^2}$$

(77)

Obviously, $[\mathbf{D}]$ is invertible as long as $\left|\int_0^x f(t)dt\right| = \sqrt{G^2 + H^2} \neq 0$. Hence, at the so-called singularities such as $x = s$ where $\left|\int_0^s f(t)dt\right| = G = H = 0$, the DTMM solution may need further attention, since $[\mathbf{D}] = [\mathbf{0}]$ and $[\mathbf{S}]$ would need singularity removal. However, as it is shown here, it is possible to show that a certain finite value for $[\mathbf{S}]$ exists when the following limit could be taken

$$[\mathbf{S}(s)] \triangleq \lim_{x \to s}[\mathbf{S}(x)] = \lim_{x \to s}[\mathbf{D}(x)]^{-1}\sinh[\mathbf{D}(x)]$$

$$= \lim_{x \to s}[\mathbf{D}(x)]^{-1}\sum_{n=0}^{\infty}\frac{[\mathbf{D}(x)]^{2n+1}}{(2n+1)!}$$

$$= [\mathbf{1}] + \lim_{x \to s} \sum_{n=1}^{\infty} \frac{[\mathbf{D}(x)]^{2n}}{(2n+1)!}$$
$$= [\mathbf{1}]$$
(78)

## APPENDIX D. INITIAL CONDITIONS

It is not difficult to show that our new DTMM solution indeed satisfies both of the two initial conditions. The new basis functions are given here as

$$\Psi_1(x) = \cos\sqrt{xG(x)}$$
$$\Psi_2(x) = x\,\text{sinc}\sqrt{xG(x)}$$
(79)

Clearly, $G(0) = 0$, and hence $\Psi_1(0) = 1$. Also, the sinc(·) function takes on the finite value of unity at zero. Hence, when multiplied by $x$, we get the second condition as $\Psi_2(0) = 0$. For the derivatives noting that $G'(x) = g(x)$ we obtain

$$\Psi_1'(x) = -\frac{G(x) + xg(x)}{2}\,\text{sinc}\sqrt{xG(x)}$$
(80)

This immediately by setting $x = 0$ implies

$$\Psi_1'(0) = 0$$
(81)

Calculation of the derivative of the second function is a bit more difficult. By direct expansion of terms we have

$$\Psi_2'(x) = \text{sinc}\sqrt{xG(x)} + x\frac{\cos\sqrt{xG(x)} - \text{sinc}\sqrt{xG(x)}}{\sqrt{xG(x)}} \times \frac{xg(x) + G(x)}{2\sqrt{xG(x)}}$$
(82)

Taylor expansion of the numerator of the first fraction and retaining the lowest-order terms gives

$$\Psi_2'(x) \sim \text{sinc}\sqrt{xG(x)} + x[xg(x) + G(x)]\frac{xG(x)\left(-\frac{1}{2!} + \frac{1}{3!}\right)}{2xG(x)}$$
$$= \text{sinc}\sqrt{xG(x)} - \frac{x}{3}[xg(x) + G(x)]$$
(83)

Now setting $x = 0$ gives the desired answer

$$\Psi_2'(0) = 1$$
(84)